# Novel Features of the Newly Discovered Field-Induced Superconducting Phase of $\lambda$-BETS$_2$FeCl$_4$

J. S. Brooks[a], L. Balicas[a], J. Matson[a], K. Storr[a], H. Kobayashi[b], H. Tanaka[b], A. Kobayashi[c], S. Uji[d], and M. Tokumoto[e]

[a] Physics/NHMFL, Florida State University, Tallahassee FL 32310 USA
[b] Institute for Molecular Science, Okazaki, Aichi 444-8585, Japan
[c] Research Centre for Spectrochemistry, Graduate School of Science, The University of Tokyo, Bunkyo-ku, Tokyo 113-0033, Japan
[d] National Institute for Materials Science, Tsukuba, Ibaraki 305-0003, Japan
[e] Nanotechnology Research Institute, National Institute of Advanced Industrial Science and Technology (AIST), Tsukuba, Ibaraki 305-8568, Japan

Tel: 1-850-644-2836; FAX: 1-850-644-5038; E-mail: brooks@magnet.fsu.edu

**Abstract**

We examine magnetic field dependent properties associated with the newly discovered field-induced superconducting state (FISC) in λ-BETS$_2$FeCl$_4$. These include the metal-to-antiferromagnetic insulator transition, the critical field of the FISC state in tilted magnetic fields, and the low-pressure, magnetic field dependence of the insulating and superconducting phases.

## I. Introduction

The organic-magnetic conductor λ-BETS$_2$FeCl$_4$ [1] undergoes a metal-to-antiferromagnetic insulator (AFI) state below 8 K. However, detailed magneto-transport and related measurements [2] demonstrated that the AFI state is removed, and the metallic state restored, above 11 T. Recently, magnetic field induced superconductivity (FISC) has been discovered in λ-BETS$_2$FeCl$_4$ at higher fields [3, 4]. Here, for magnetic fields carefully aligned parallel to the conducting layers of this quasi-two dimensional material, a novel superconducting state is stabilized above about 18 T for temperatures below 5 K. It has since been shown that this FISC state is re-entrant to the normal metallic state above 42 T[5], and that the FISC state moves to lower magnetic fields with increasing Ga concentration in the alloy compound λ-BETS$_2$Fe$_x$Ga$_{1-x}$Cl$_4$ [6]. A field cancellation mechanism where the π – d exchange field in the conducting layers is removed at high magnetic field is proposed as the main mechanism that allows the high field superconducting state (i.e. the Jaccarino-Peter effect as observed in the Chevrel materials[7, 8]), and some theoretical treatments have appeared along these lines[5, 9]. The purpose of this short paper is to consider the nature of the magnetic field-dependent

AFI transition, the anisotropic nature of the FISC state, and the anomalous behavior of the superconducting, AFI, and FISC states at very low pressures.

## II. Resistive transition

The onset of the anti-ferromagnetic transition at 8 K is accompanied by a dramatic change in the resistivity of the material, which can increase by over 6 orders of magnitude. The nature of this transition is unique, since two mechanisms may co-operate to stabilize the AFI state, due to the π-d interaction. The columnar nature of the BETS molecular structure alone is susceptible to 1-D instability (i.e. a SDW transition), but the cooperative AF interaction of the $Fe^{3+}$ anions enhances this effect, making the transition essentially first order. The theoretical question arises[9] as to the Mott insulator vs. band (nesting) behavior of this transition. This behavior is shown in Fig. 1a for one of four samples where the M-AFI transition was studied vs. magnetic field up to 8 T. In these measurements care was taken to insure that the connecting leads in the four-terminal measurements had an inter-connecting resistance of 50 G ohms. An electrometer with 200 T ohm input impedance was used to measure the voltage, and currents in the range 1 to 10 nA insured ohmic behavior over the entire temperature range. The logarithmic derivative (which gives the thermal activation energy $E_a$ vs. temperature) is shown in Fig. 1b. Although there is considerable scatter in the data, it is generally observed (in all samples investigated to date) that the activation gap derived is of order 100 K at $T_{AFI}(B=0)$, but decreases to about 20 K for $T = T_{AFI}(B=0)/2$. Fig. 1c shows the magnetic field dependence of $T_{AFI}$ and $E_a$ (at 4.5 K) over the range investigated. Of note is that the activation energy (e.g. for the zero field data), when compared with the standard BCS relationship $2E_a/T_c = 3.52$, shows a significant deviation towards higher values, with $2E_a/T_c$ approaching 25 at $T_{AFI}(B=0)$. The theoretical prediction of Cepas et al. is that, if the AFI is a Mott insulator, $E_a$ should not vary significantly with field[9]. In Fig. 1c, $E_a$ does not appear to decrease with increasing field, but rather, there is some indication for an increase in $E_a$ with field. Experimental uncertainties exclude any quantitative comparison at this time, and further work is needed to determine the field dependence of $E_a$.

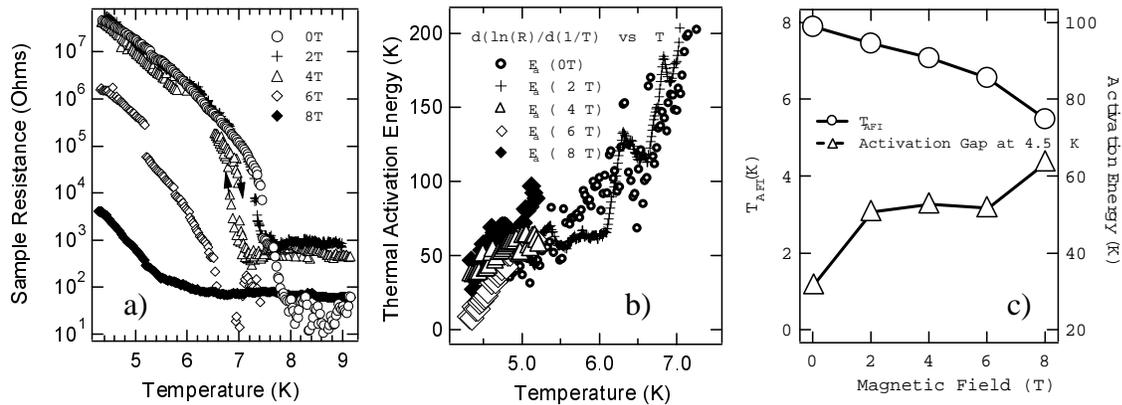

Figure 1. Resistance vs. temperature for $\lambda$-BETS$_2$FeCl$_4$ for different magnetic fields (a), corresponding logarithmic derivative plot for the asymptotic low temperature behavior (b), and field dependence of $T_{AFI}$ and $E_a$(at 4.5 K) vs. field (c).

## III. Magnetoresistive anomalies in the normal and FISC phases

Data taken early in our the investigation of the FISC state for tilted magnetic fields is shown in Fig. 2a, where at 35 mK, the FISC was measured vs. incremental changes in the magnetic field orientation. We emphasize that this data is for θ > 60 degrees, well past the angle where Shubnikov - de Haas ( SdH ) oscillations, due to the 2D Fermi surface, are apparent[4]. There are a number of resistive anomalies, shown in Fig. 2b that appear above 11 T where the normal metallic state appears (inset), and also as the FISC phase is stabilized at higher fields for θ -> $90^0$ (steps shown as dashed lines). Analysis of the inverse field dependence of these anomalies indicates that they are not associated with the SdH effect, nor any sort of quantized resistance or conductivity behavior. Since these features appear to be dependent on sample and also on the axis of rotation, we speculate that their origin is most likely due to magnetic domains in the $FeCl_4$ structure. In this scenario, domain formation (which would be highly dependent on field direction and sample characteristics) would influence the conductivity and superconductivity in the conducting BETS layers. This is a further example of the highly unique nature of the interaction of the d and π electron systems.

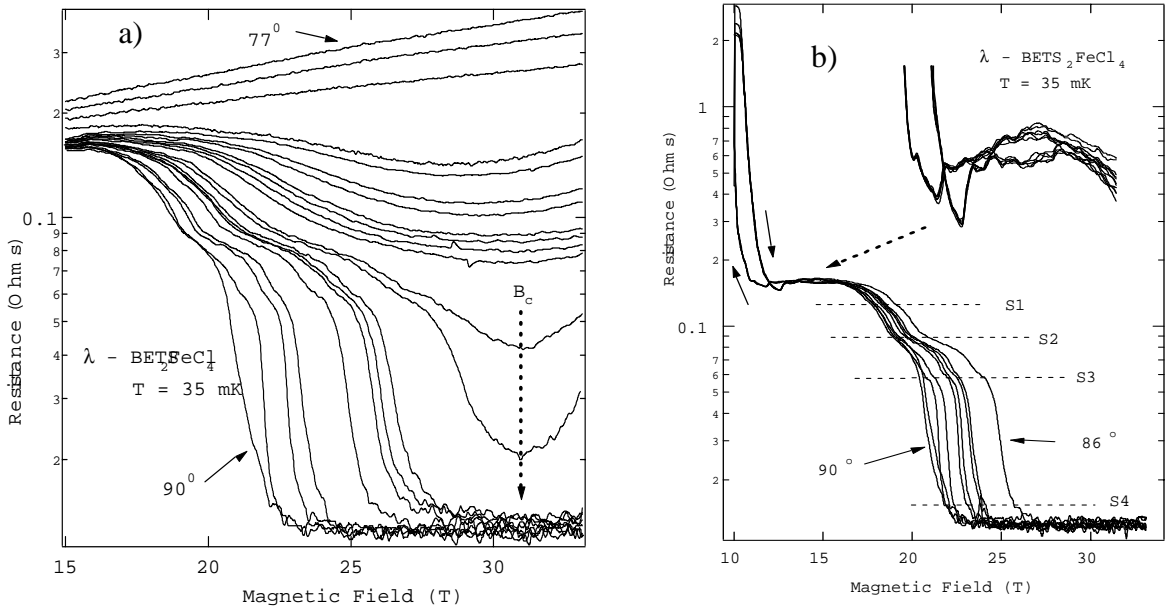

Figure 2. a) Magnetoresistance of λ-$BETS_2FeCl_4$ for steps (of order 1degree) in field rotation in the a-b plane. The center of the FISC state is defined by $B_c$. b) Detail of a) showing up and down field sweeps (solid arrows) in the range θ = 86 to 90 degrees in approximately 0.5 degree steps. Dashed lines: position of resistance steps. Inset, detail of AFI-to-metallic region.

## IV. Examination of 2D superconducting behavior

We have examined the angular dependence of the lower critical field regime in the FISC state in Figure 2. There are three important parameters necessary to stabilize the FISC in the Jaccarino-Peter model. First are the in-plane (Pauli limit) and inter-plane (orbital limit) critical fields for the pure $GaCl_4$ system, 12 T and 3.5 T respectively[10, 11], and the third is the exchange field which may be estimated as $B_J \approx B_c$. For the parallel applied field $B_{//} = B_c$, the internal field $B_{int}$ is essentially zero, and we may therefore measure the magnitude of the internal field as increasing in <u>both</u> directions away from this point, maintaining a superconducting state until a difference of 12 T, the upper (Pauli limit) critical field $B_p$ of the non-magnetic $\lambda$-BETS$_2$GaCl$_4$ [11, 12] is reached. This situation has some interesting consequences, since the superconducting state does not exist until the exchange field is reduced below the value 12 T for in-plane field, and as our previous analysis shows (see Ref. [5]), for inter-plane fields below 3.5 T. We may therefore roughly consider the loci of external magnetic field and in-plane orientation ($\theta$ taken as the direction of field with respect to the conducting planes) where superconductivity is present as follows

$B_{//} = B_{tot}\sin(\theta) > B_J - B_p$ ; $B_\perp = B_{tot}\cos(\theta) < 3.5$ T ; $B_{//} = B_{tot}\sin(\theta) < B_J + B_p$

This argument, which describes the loci of field magnitude and orientation where the FISC can exist, is given schematically in Fig. 3a. It leads to the prediction of a critical value of B and $\theta$ for the onset of the FISC.

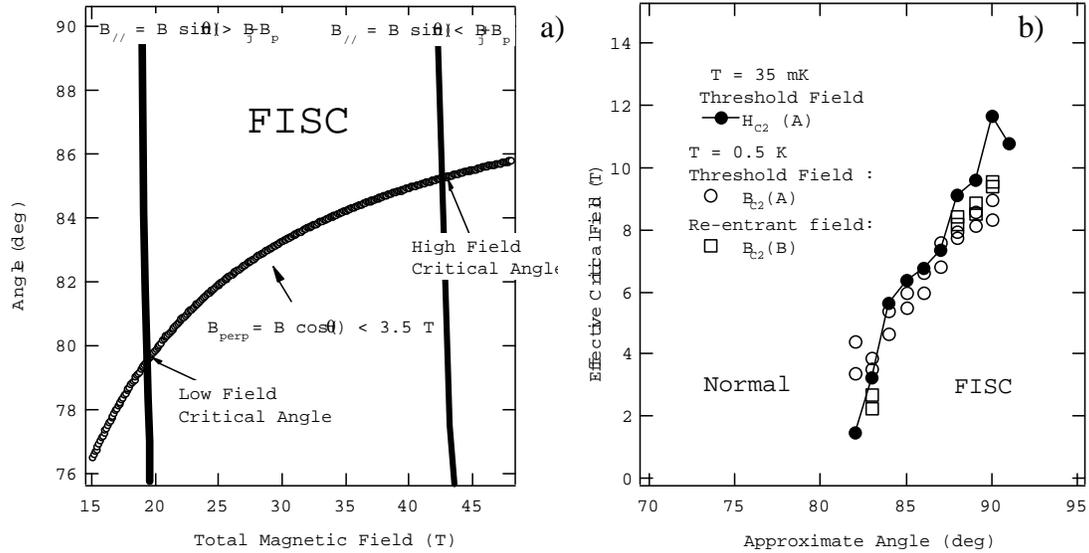

Figure 3. a) Results of the simple model for the loci of field orientation and magnetic field within which the Jaccarino-Peter, orbital, and Pauli limit requirements are met for superconductivity. b) The effective critical field taken for the 10% criteria for the resistive transition associated with the FISC state.

It is important to consider, based on the Jaccarino Peter mechanism[7, 8], the nature of the critical field in the FISC state. The relation of in-plane, inter-plane, and exchange fields may be further pursued by considering the results in Fig. 3b, as derived from Fig. 2a and also the re-entrant data from Ref. [5]. Unlike traditional 2D superconductivity, where anisotropic effective mass or 2D layer models (dependent only on $\theta$, $B_\perp$, and $B_{//}$) may be used to describe the upper critical field[13], we find here an additional parameter, the exchange field cancellation term. This leads to a unique situation when we consider the angular dependence of the critical field: $B_{c2}(A)$ is defined as the critical field at the onset of the FISC state below $B_c$, and $B_{c2}(B)$ is defined as the critical field at the re-entrant boundary above $B_c$ (see Ref. [5]). We show in Fig. 3b the onset of superconductivity in terms of the effective internal field vs. angle, based on a simple 10% $R_N$ criterion, for $B_{c2}(A)$ (with <u>decreasing</u> total field) and $B_{c2}(B)$ (for <u>increasing</u> total field) as θ approaches 90 degrees. The data near 90 degrees is "cusp-like", approximating similar behavior for $B_{c2}$ in the non-magnetic superconductor λ-$BETS_2GaCl_4$[12]. However, away from 90 degrees, the critical field deviates from standard 2D behavior, and the FISC state disappears below 70 degrees due to the arguments represented in Fig. 3a.

## V. Anomalous dependence of SC, AFI, and FISC states on very low pressure

We have carried out preliminary, low-pressure measurements of λ-$BETS_2FeCl_4$ where the magnetic field was aligned along the c-axis. The results are a bit surprising, given previous work where, at zero magnetic field and for a pressure of 3 kbar, the AFI state was removed in favor of superconductivity[14]. Our results are shown in Fig. 4, where at the lowest (ε) pressure studied, a standard clamp cell with Flourinert was fixed only "finger tight". Since it is well known that there is considerable pressure loss when clamps of this type are cooled down, the actual value of pressure and/or irregular strain due to freezing of the fluid is at present unknown. Nevertheless, the results are striking, since it is evident that a superconducting phase region is stabilized in the temperature range between 6 K and 4 K at low fields. At higher fields, the AFI, and then the metallic states are stabilized, and eventually, the FISC phase appears. A tentative phase diagram of this behavior is shown in Fig. 4c. Further work is in progress in the low pressure regime to explore this complex and unexpected pressure dependent behavior.

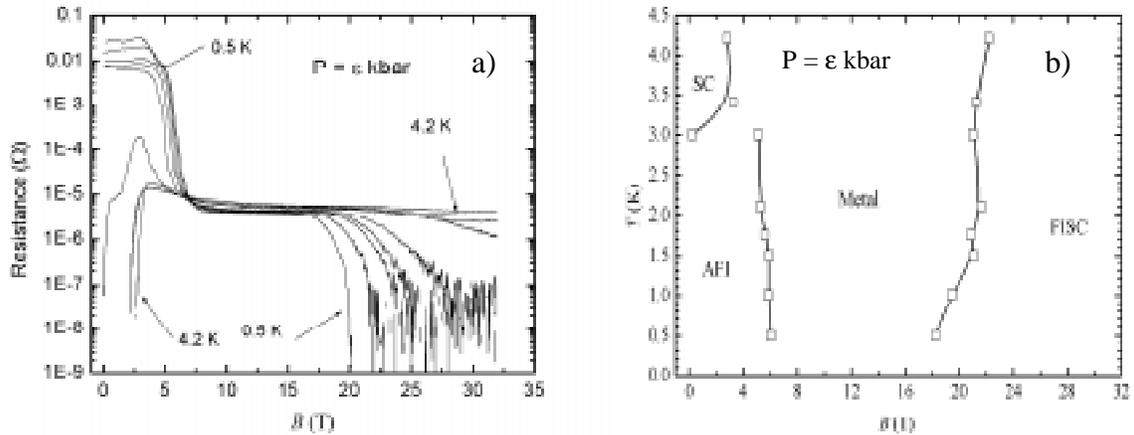

Figure 4. Pressure dependent behavior of the magnetoresistance and for λ-$BETS_2FeCl4$ with the magnetic field aligned in-plane. a) Behavior where the clamp cell is "finger tight". The temperature is decreased from 4 K to 0.5 K. Note that the sample is superconducting at 4 K at zero field, but goes into the AFI state at lower temperatures or at higher fields. The FISC state has the opposite temperature dependence. b) An approximate phase diagram of the low pressure behavior.

Acknowledgements:
In light of the tragedy that occurred during this conference, this work is submitted in consideration of better international communication, collaboration, and friendship in the spirit of ISCOM. This work is supported by NSF-DMR 99-71474.